\documentclass[11pt]{article}
\usepackage[margin=1in]{geometry}
\usepackage{amsmath,amssymb,amsthm}
\PassOptionsToPackage{hyphens}{url}
\usepackage[hidelinks]{hyperref}

\newtheorem{theorem}{Theorem}[section]
\newtheorem{lemma}[theorem]{Lemma}
\newtheorem{proposition}[theorem]{Proposition}
\newtheorem{corollary}[theorem]{Corollary}
\theoremstyle{definition}
\newtheorem{definition}[theorem]{Definition}
\newtheorem{remark}[theorem]{Remark}

\newcommand{\per}{\operatorname{per}}
\newcommand{\Pf}{\operatorname{Pf}}
\newcommand{\sgn}{\operatorname{sgn}}
\newcommand{\cyc}{\operatorname{cyc}}
\newcommand{\adj}{\operatorname{adj}}
\newcommand{\tr}{\operatorname{tr}}
\newcommand{\Mat}{\operatorname{Mat}}
\newcommand{\Hom}{\operatorname{Hom}}
\newcommand{\Hilb}{\operatorname{Hilb}}
\newcommand{\Spec}{\operatorname{Spec}}
\newcommand{\length}{\operatorname{length}}
\newcommand{\dc}{\operatorname{dc}}
\newcommand{\pc}{\operatorname{pc}}
\newcommand{\bdc}{\overline{\operatorname{dc}}}
\newcommand{\bpc}{\overline{\operatorname{pc}}}
\newcommand{\PP}{\mathbb{P}}
\newcommand{\A}{\mathbb{A}}
\newcommand{\C}{\mathbb{C}}
\newcommand{\Q}{\mathbb{Q}}
\newcommand{\Z}{\mathbb{Z}}
\newcommand{\LL}{\mathbb{L}}
\newcommand{\OO}{\mathcal{O}}
\newcommand{\cZ}{\mathcal{Z}}
\newcommand{\ii}{\hat\imath}
\newcommand{\jj}{\hat\jmath}

\title{Cubic lower bounds for the permanent in odd characteristic,\\ for Pfaffian representations, and for polynomial entries}
\author{Karthik Sheshadri}
\date{October 28, 2026}

\begin{document}
\maketitle

\begin{abstract}
The determinantal complexity of a polynomial is the order of the smallest matrix whose entries are polynomials of degree at most one and whose determinant equals the polynomial. A lower bound for the permanent in this measure is the algebraic form of the question whether the permanent is harder to compute than the determinant. Over the complex numbers, the best known lower bound grows cubically in the size of the permanent. It holds even when the permanent only has to be a limit of such determinants.

I prove that the cubic lower bound holds over every algebraically closed field of odd characteristic $p$, with a constant that depends on $p$. Hence the cubic bound does not depend on characteristic zero. I also prove the bound for Pfaffians, which are square roots of determinants of skew symmetric matrices. Lower bounds for Pfaffians imply the same bounds for determinants. Further, I prove the bound, divided by $a$, for matrices whose entries are polynomials of degree at most $a$. All three results come from one lower bound for smooth hypersurfaces. I prove it in every characteristic by counting intersection points together with their multiplicities.
\end{abstract}

\section{Introduction}

The permanent of an $n\times n$ matrix $Y$ is
\[
\per_n(Y)=\sum_{\sigma}\prod_{i=1}^{n}Y_{i,\sigma(i)},
\]
where $\sigma$ runs over the permutations of $\{1,\dots,n\}$. The determinant is the same sum with the sign of $\sigma$ attached to each term. The determinant can be computed quickly by Gaussian elimination. However, no fast method is known for the permanent. Valiant~\cite{valiant} gives this gap an algebraic form. He shows that a polynomial computed by a formula of size $s$ is the determinant of a matrix of order linear in $s$ whose entries are affine linear forms, i.e., polynomials of degree at most one. He also shows that, in characteristic other than two, the permanent is complete under such substitutions for the class VNP, an algebraic analogue of NP.

A determinantal representation of a polynomial $F(x_1,\dots,x_N)$ is a square matrix $M(x)$ whose entries are affine linear forms in $x$ and whose determinant is $F$. The determinantal complexity $\dc(F)$ is the smallest order of such a matrix. Because of Valiant's results, the question whether the permanent is harder than the determinant becomes the question whether $\dc(\per_n)$ grows faster than every polynomial in $n$. The border determinantal complexity $\bdc(F)$ is the smallest $m$ such that $F$ is a limit of determinants of $m\times m$ matrices with affine linear entries. Here the limit is taken coefficient by coefficient. Every representation is also such a limit. Hence $\bdc(F)\le\dc(F)$, and a lower bound on $\bdc(F)$ is the stronger statement.

Mignon and Ressayre~\cite{mr} show that $\dc(\per_n)\ge n^2/2$ in characteristic zero. Cai, Chen and Li~\cite{ccl} extend this quadratic bound to every characteristic other than two. In characteristic two the permanent equals the determinant, so no bound beyond $n$ is possible there. A manuscript from September 2026~\cite{oai} gives the first cubic bound, $\bdc_{\C}(\per_n)=\Omega(n^3)$. Its proof has two parts. Firstly, it restricts the permanent to a linear subspace on which the permanent becomes a smooth hypersurface of degree about $n/3$ in about $n^2/100$ variables. Secondly, it shows that a smooth hypersurface of degree $d$ in $N$ variables has border determinantal complexity at least $(N-1)(d-1)/(4e)$, where $e$ is Euler's number.

Both parts of that proof use characteristic zero. The first part selects one coefficient inside a permanent, and it uses factorials as weights to do so. In characteristic $p$, every factorial $j!$ with $j\ge p$ is zero, so this selection breaks down. The second part counts the points where the hypersurface meets several of its first derivatives, and it assumes that these points are simple. In positive characteristic this can fail. For example, in characteristic $p$ every first derivative of $x_1^{p+1}+\dots+x_N^{p+1}$ is the $p$th power of a linear form. Hence every intersection point is repeated.

Two other models of representation are natural. The first uses Pfaffians. A square matrix $M$ is alternating if $M^T=-M$ and its diagonal is zero. Its Pfaffian $\Pf M$ is a polynomial in the entries whose square is $\det M$. A Pfaffian representation of $F$ of half order $m$ is an alternating matrix of order $2m$ with affine linear entries whose Pfaffian is $F$. Every determinantal representation of order $m$ gives a Pfaffian representation of half order $m$, by~\eqref{eq:pfdet} below. Therefore a lower bound for Pfaffians is also a lower bound for determinants. The second model allows entries that are polynomials of degree at most $a$. It lies between Valiant's model and the trivial representation $[F]$, which has a single entry. I write $\bpc(F)$ for the border complexity of Pfaffian representations, and $\bdc_a(F)$ and $\bpc_a(F)$ for the border complexities with entries of degree at most $a$. To the best of my knowledge, nothing in the published literature addresses the cubic bound in positive characteristic, for Pfaffians, or for polynomial entries.

My contributions are three cubic lower bounds. Firstly, the cubic bound holds over every algebraically closed field of odd characteristic $p$, with a constant that depends on $p$. Secondly, it holds for Pfaffian representations, both over $\C$ and in odd characteristic. Thirdly, it holds for representations whose entries have degree at most $a$, divided by $a$. Note that the order of the permanent in the first result matters. Valiant's completeness theorem already makes a determinant of order $k$ a restriction of a permanent of order polynomial in $k$. However, a permanent of order $k^c$ would only give a bound of order $n^{3/c}$. My construction has order linear in $k$ for each fixed $p$. Note also that the third result does not follow from the first. An entry of degree two can have many terms. For example, the $1\times1$ matrix $[x_1^2+\dots+x_N^2]$ has one entry of degree two. However, every representation of $x_1^2+\dots+x_N^2$ with affine entries has order at least $(N-1)/(4e)$ in characteristic other than two, by Theorem~\ref{thm:smooth}.

The precise statements use the constant $B_p=5p+38$.

\begin{theorem}\label{thm:main}
Let $p$ be an odd prime and $K$ an algebraically closed field of characteristic $p$. For every $n\ge 128B_p$,
\[
\bdc_K(\per_n)\ge\frac{n^3}{270\,e\,B_p^3},\qquad \bpc_K(\per_n)\ge\frac{n^3}{540\,e\,B_p^3}.
\]
Over the complex numbers, $\bpc_{\C}(\per_n)\ge n^3/(8.8\cdot 10^{6}\,e)$ for every $n\ge 1408$. The same bounds hold for the exact complexities.
\end{theorem}

\begin{theorem}\label{thm:poly}
Let $k=\lfloor n/11\rfloor$ over $\C$ and $k=\lfloor n/B_p\rfloor$ in odd characteristic $p$, and suppose $k\ge 128$. For every integer $1\le a\le k/6$, over $\C$,
\[
\bdc_a(\per_n)\ge\frac{\log(4/3)}{9600}\cdot\frac{k^3}{a},\qquad \bpc_a(\per_n)\ge\frac{\log(4/3)}{19200}\cdot\frac{k^3}{a},
\]
and in characteristic $p$ the same bounds hold with $1008$ and $2016$ in place of $9600$ and $19200$. Both are therefore $\Omega(n^3/a)$ over $\C$ for $a\le n/72$, and $\Omega_p(n^3/a)$ in characteristic $p$ for $a\le n/(12B_p)$.
\end{theorem}

The constants are poor and depend on $p$. Note also that some bound on $a$ in Theorem~\ref{thm:poly} is necessary. Indeed, the $1\times 1$ matrix $[\per_n]$ is a representation with one entry of degree $n$.

All three results rest on one lower bound for smooth hypersurfaces, Theorem~\ref{thm:smooth}, which holds in every characteristic. To prove it, I intersect the hypersurface with $N-2$ general first derivatives of its equation. The intersection is a finite set of points with multiplicities, and the sum of the multiplicities is $d(d-1)^{N-2}$. I follow this sum through the limit. I then bound it from above by a B\'ezout count on a larger space, which also records the kernel vectors of the matrix. Since the count uses multiplicities throughout, repeated points do no harm.

For Pfaffians I do not need a second argument. Let $M$ be alternating of order $2m$, and let $E$ be its submatrix on the rows and columns $3,\dots,2m$. Let $M'$ be $M$ with its first row and second column deleted. Then $\det M'=-\Pf(E)\Pf(M)$ (Lemma~\ref{lem:pfdet}). After a constant change of coordinates, $\Pf E$ is nonzero at every point of the count. Hence, near these points, $\Pf M$ is a nonzero multiple of a determinant of order $2m-1$, and the count for determinants applies.

For the permanent in characteristic $p$, I need two more ingredients. Firstly, the smooth hypersurface is a general linear section of $G_D(X)=[s^qt^q]\det(X+sI+tD)$, the coefficient of $s^qt^q$ in a determinant. This section is smooth because the set of matrices $X$ where every derivative of $G_D$ vanishes is small. Theorem~\ref{thm:crit} proves this in every characteristic. Secondly, $G_D$ must be a restriction of a permanent of order linear in $k$. Recall that $p-1=-1$ in characteristic $p$. Hence, if a permanent gives each cycle of a permutation one of $p-1$ colours, it counts the permutation $(-1)^c$ times, where $c$ is its number of cycles. Up to sign, this is the determinant. A permanent that forces exactly $q$ of $L$ marked entries to be used then extracts the coefficient (Theorem~\ref{thm:detperm}).

The remainder of this paper is organized as follows. Section~\ref{sec:reps} fixes the representations. Section~\ref{sec:ingredients} states the two geometric results that the proof needs. Section~\ref{sec:perm} builds the permanents in characteristic $p$, and Section~\ref{sec:proofs} proves Theorems~\ref{thm:main} and~\ref{thm:poly}. Section~\ref{sec:ai} describes how I used artificial intelligence. Appendices~\ref{app:smooth} and~\ref{sec:crit} prove the two geometric results, and Appendix~\ref{app:selectors} builds the selectors. A reader who accepts the geometric results can skip the appendices.

\section{Representations, limits and Pfaffians}\label{sec:reps}

In this section, I fix the matrices that are allowed and say what it means for a polynomial to be a limit of their determinants or Pfaffians.

\paragraph{Notation.} Throughout, $K$ is an algebraically closed field of any characteristic, unless I say otherwise, and $e=\exp(1)$. For a form $F$ in $N$ variables, $V(F)\subset\PP^{N-1}$ is its zero scheme, i.e., its zero set together with multiplicities. For $w\in K^N$, the first polar $\partial_wF=\sum_j w_j\,\partial F/\partial x_j$ is the derivative of $F$ in the direction $w$. A bracket such as $[s^it^j]P$ denotes the coefficient of $s^it^j$ in $P$. A zero dimensional scheme is a finite set of points, each with a finite dimensional local ring. Its length is the dimension of its coordinate ring over $K$, and the length of a point is the dimension of its local ring. For a set of simple points, the length is just the number of points. Finally, $e_r$ is the elementary symmetric polynomial of degree $r$.

\begin{definition}\label{def:classes}
Fix $a\ge 1$. Let $D_{m,a}$ be the set of polynomials $\det M(x)$, where $M$ is an $m\times m$ matrix whose entries are polynomials in $x_1,\dots,x_N$ of degree at most $a$. Let $P_{2m,a}$ be the set of polynomials $\Pf M(x)$, where $M$ is alternating of order $2m$ with entries of the same kind. Both sets lie in the space of polynomials of degree at most $am$, and I take Zariski closures there. Put
\[
\bdc_a(F)=\min\{m: F\in\overline{D_{m,a}}\},\qquad \bpc_a(F)=\min\{m:F\in\overline{P_{2m,a}}\}.
\]
The exact complexities $\dc_a$ and $\pc_a$ are defined in the same way without the closure. For $a=1$ I drop the subscript.
\end{definition}

Three simple facts about these classes are used later. Firstly, an affine substitution $x\mapsto\Lambda x+x_0$ keeps entry degrees at most $a$. Secondly, a nonzero scalar can be absorbed into one row of a determinant, or into an alternating matrix by a constant congruence. Thirdly, for every $m\times m$ matrix $M$ over a commutative ring,
\begin{equation}\label{eq:pfdet}
\Pf\begin{pmatrix}0&M\\-M^T&0\end{pmatrix}=(-1)^{m(m-1)/2}\det M.
\end{equation}
Hence all four complexities can only decrease under affine restriction, and $\bpc_a\le\bdc_a$. Note also that $\per_r$ is an affine restriction of $\per_n$ for $r\le n$. To see this, put an identity block in one corner.

\section{The two geometric ingredients}\label{sec:ingredients}

The proof needs two results from algebraic geometry. The first is a lower bound for smooth hypersurfaces. The second gives a smooth hypersurface with many variables and large degree. I state both here, so that the rest of the paper can be read without their proofs. Appendix~\ref{app:smooth} proves the first, and Appendix~\ref{sec:crit} proves the second.

\subsection{A lower bound for smooth hypersurfaces}

Let $F$ be a form of degree $d\ge2$ in $N\ge2$ variables. I say that $V(F)$ is smooth if the partial derivatives of $F$ have no common zero on $V(F)$. The proof of the next theorem is in Appendix~\ref{app:smooth}. The idea is as follows. Firstly, the hypersurface meets $N-2$ general first polars in a finite scheme of length $d(d-1)^{N-2}$. Secondly, suppose that $F$ is a determinant. At each point of this scheme the matrix loses rank, and the vanishing of a polar there becomes a condition on the left and right kernel vectors of the matrix. Thirdly, I add the kernel vectors as new coordinates. The scheme then becomes a set of isolated points of a system of equations of low degree. A B\'ezout count bounds the total length of these points by a coefficient that grows only polynomially with the order of the matrix.

\begin{theorem}\label{thm:smooth}
Let $F$ be a form of degree $d\ge2$ in $N\ge2$ variables over an algebraically closed field of any characteristic, with $V(F)$ smooth.
\begin{enumerate}
\item[(i)] If $\bdc(F)\le m$, then $d(d-1)^{N-2}\le 2^{N-2}\binom{2m-1}{N-1}$, and if $\bpc(F)\le m$, the same holds with $4m-3$ in place of $2m-1$. Consequently
\[
\bdc(F)\ge\frac{(N-1)(d-1)}{4e}+\frac12,\qquad \bpc(F)\ge\frac{(N-1)(d-1)}{8e}+\frac34.
\]
\item[(ii)] If $0<c<1$ and $a\le cd$, then with $L_c=\log\bigl(2/(1+c)\bigr)$,
\[
\bdc_a(F)\ge\frac{(N-1)(1-c)L_c}{8}\cdot\frac da+\frac12,\qquad \bpc_a(F)\ge\frac{(N-1)(1-c)L_c}{16}\cdot\frac da+\frac34.
\]
\end{enumerate}
In addition, $\bdc_a(F)$ and $\bpc_a(F)$ are at least $\lceil d/a\rceil$.
\end{theorem}

Proposition~\ref{prop:fermat} in Appendix~\ref{app:smooth} shows that these bounds have the right order of growth.

\subsection{A smooth hypersurface of large degree}

Theorem~\ref{thm:smooth} applies to any smooth hypersurface. To reach the permanent, I need one particular smooth hypersurface with large $Nd$ that is a restriction of a small permanent. It comes from the polynomial $G_D$ below, restricted to a general linear subspace. Such a restriction is smooth when the critical set of $G_D$ is small. Here the critical set is the set of points at which every partial derivative of $G_D$ vanishes. Theorem~\ref{thm:crit} below bounds the dimension of the critical set in every characteristic. Its proof in Appendix~\ref{sec:crit} follows the determinant side of~\cite{oai}. The only outside input is a dimension bound for plane curves of given genus~\cite{cht}.

Fix $k\ge2$ and put
\[
q=\Bigl\lfloor\frac{k-1}{3}\Bigr\rfloor,\qquad d=k-2q,\qquad c=d-1,\qquad T=\binom{q+2}{2},
\]
so that $2q+c=k-1$ and $c\ge q$. I choose distinct $\theta_1,\dots,\theta_k\in K$ and let $D=\mathrm{diag}(\theta_1,\dots,\theta_k)$. Define
\[
G_D(X)=[s^qt^q]\det(X+sI_k+tD),\qquad \Gamma_D=\{X\in\Mat_k(K):\nabla G_D(X)=0\},
\]
where the gradient is taken with respect to all $k^2$ entries of $X$.

\begin{theorem}\label{thm:crit}
In every characteristic, $\dim\Gamma_D\le k^2+k-T$.
\end{theorem}

The idea of the proof is as follows. To each matrix $X$ I attach a plane curve $C_X$ of degree $k$. Its normalisation defect $\delta(C_X)$ measures its singularities. It is the length of $\nu_*\OO_{\widetilde C}/\OO_C$, where $\nu:\widetilde C\to C$ is the normalisation, i.e., the smooth curve that maps onto $C$ and is an isomorphism over the smooth points of $C$. For example, a node contributes one to $\delta$. Firstly, if $X\in\Gamma_D$, then $\delta(C_X)\ge T$. Secondly, the curves of degree $k$ with $\delta\ge T$ form a set of codimension at least $T$. Thirdly, the matrices with a given curve form a set of dimension at most $k(k-1)/2$.

I now use Theorem~\ref{thm:crit} to obtain the smooth hypersurface. Since $\Gamma_D$ has codimension at least $T-k$, a general linear section can have up to $T-k$ variables and still be smooth. I take all of them.

\begin{corollary}\label{cor:smooth}
Let $k\ge128$, let $q$, $d$ and $T$ be as above, and put $N=T-k$. For a general linear map $\Lambda:K^N\to\Mat_k(K)$, the form $F=G_D\circ\Lambda$ is nonzero of degree $d$, and $V(F)\subset\PP^{N-1}$ is smooth. Moreover
\begin{equation}\label{eq:params}
N-1\ge\frac{k^2}{21},\qquad d-1\ge\frac{k-1}{3},\qquad d\ge\frac k3.
\end{equation}
\end{corollary}

Appendix~\ref{sec:crit} proves Theorem~\ref{thm:crit} and Corollary~\ref{cor:smooth}.

\section{Permanents in odd characteristic}\label{sec:perm}

Corollary~\ref{cor:smooth} gives a smooth restriction of $G_D$. However, $G_D$ is not yet a restriction of a permanent. Over $\C$, the manuscript~\cite{oai} selects the coefficient of $s^qt^q$ inside a permanent, using factorials as weights. In characteristic $p$ this fails once $p\le q$. In this section, I replace that step by two constructions. The first is a permanent of linear order that forces exactly $q$ of $L$ marked entries to be used. The second writes the determinant of order $k$ as a permanent of order $(5p-4)k$. From now on, $p$ is an odd prime and $K$ has characteristic $p$.

I read a matrix as a weighted bipartite graph between its rows and its columns. A perfect matching pairs every row with a distinct column, and its weight is the product of the entries that it uses. The permanent is the sum of the weights of all perfect matchings. A marked entry is an entry that I single out, and its selection bit records whether a given matching uses it. If a marked entry is one summand of an entry, only that summand is meant. All constants below lie in $\Z[1/2]$. Hence every identity below holds in every odd characteristic.

Both constructions need a way to force two marked entries to be used together or not at all. The equality gadget does this. It adds three new rows and three new columns, and Appendix~\ref{app:selectors} gives it.

\subsection{A selector of linear order}

I now state the selector. It must check that exactly $q$ of $L$ bits are equal to one. I do this the way one adds numbers in binary. A circuit of full adders adds the bits, and its output is fixed to the binary digits of $q$. Appendix~\ref{app:selectors} builds the circuit and proves the following proposition.

\begin{proposition}\label{prop:selector}
For $L\ge1$ and $0\le q\le L$ there is a matrix $E_{L,q}(z)$ of order at most $17L$, with entries in $\Z[1/2]\cup\{z_1,\dots,z_L\}$, such that $\per E_{L,q}=e_q(z_1,\dots,z_L)$. Each $z_i$ occurs in one entry, and no row or column contains two of them.
\end{proposition}

\begin{remark}\label{rem:select}
Let $e_1,\dots,e_L$ be marked entries of weights $w_i$ in some matrix. Couple each $e_i$ by an equality gadget to the $z_i$ entry of a disjoint copy of $E_{L,q}$, with $z_i$ set to $1$. The permanent of the result is the sum of the weights of the matchings of the original matrix that use exactly $q$ of the $e_i$. The order grows by at most $20L$.
\end{remark}

\subsection{The determinant as a permanent}

I now write the determinant as a permanent. Recall that $p-1=-1$ in characteristic $p$. Suppose that each cycle of a permutation $\sigma\in S_k$ receives one of $r=p-1$ colours. Then $\sigma$ is counted $r^{\cyc(\sigma)}=(-1)^{\cyc(\sigma)}$ times, where $\cyc(\sigma)$ is the number of cycles of $\sigma$. Since $(-1)^{\cyc(\sigma)}=(-1)^k\sgn\sigma$, this count gives the determinant up to sign. I realise the colours by $r$ copies of the matrix, and a small gate makes each index available in exactly one copy. The count gate is
\[
C_r(z)=\begin{pmatrix}\mathrm{diag}(z_1,\dots,z_r)&\mathbf 1\\\mathbf 1^T&0\end{pmatrix},\qquad \per C_r(z)=e_{r-1}(z).
\]
To see the formula, note that a perfect matching leaves out exactly one diagonal entry, say the $j$th. It is then completed in one way, through the last row to column $j$ and from row $j$ to the last column.

\begin{theorem}\label{thm:detperm}
Let $p$ be an odd prime and $r=p-1$.
\begin{enumerate}
\item[(i)] For every $k\times k$ matrix $A$ over a commutative $K$-algebra there is a matrix $B(A)$ of order $(5p-4)k$, whose entries are entries of $A$ or constants, with $\per B(A)=(-1)^k\det A$.
\item[(ii)] For $q=\lfloor(k-1)/3\rfloor$, the polynomial $(-1)^kG_D(X)$ is the permanent of a matrix of order at most $(5p+38)k$ whose entries are entries of $X$ or constants. In particular, $G_D$ is an affine restriction of $\per_{n'}$ for some $n'\le(5p+38)k$.
\end{enumerate}
\end{theorem}

\begin{proof}
For (i), I take $r$ disjoint copies of $A$, with rows and columns $(c,i)$ in copy $c$. I add to each diagonal entry $(c,i)$ a marked summand $1$, called the bypass. For each index $i$, I add a count gate, and I couple the bypass at $(c,i)$ to the $c$th diagonal entry of that gate by an equality gadget, for $c=1,\dots,r$. The gate makes exactly $r-1$ of the bypasses at $i$ be used. Hence exactly one copy has the index $i$ free for the entries of $A$. Let $J_c$ be the set of indices that are free in copy $c$. Then $(J_1,\dots,J_r)$ runs over the ordered partitions of $\{1,\dots,k\}$, with empty parts allowed, and
\[
\per B(A)=\sum_{J_1\sqcup\dots\sqcup J_r=\{1,\dots,k\}}\ \prod_{c=1}^r\per A_{J_c,J_c}=\sum_{\sigma\in S_k}r^{\cyc(\sigma)}\prod_iA_{i,\sigma(i)}.
\]
The second equality holds because a term on the left is a permutation together with a colour for each of its cycles. In characteristic $p$ the right side is $(-1)^k\det A$. The order is $rk$ for the copies, $(r+1)k$ for the gates and $3rk$ for the gadgets, which is $(5r+1)k$ in all.

For (ii), I apply (i) to $X$. For each $i$, I add two vertices $s_i$ and $t_i$, each with its own row and column, carrying marked loops of weights $z_i$ and $w_i$. For every copy $c$, I add the entries $1$ from $(c,i)$ to $s_i$ and back, $\theta_i$ from $(c,i)$ to $t_i$, and $1$ from $t_i$ to $(c,i)$. A new vertex can attach only to the copy in which $i$ is free. There it acts as an extra diagonal entry, equal to $1$ for $s_i$ and to $\theta_i$ for $t_i$. Otherwise it uses its loop. Note that $s_i$ and $t_i$ cannot both attach at $i$. With all $z_i=z$ and all $w_i=w$, the colouring count of (i) gives
\[
H(X,z,w)=(-1)^kz^kw^k\det(X+z^{-1}I+w^{-1}D),\qquad [z^{k-q}w^{k-q}]H=(-1)^kG_D(X).
\]
Remark~\ref{rem:select}, applied to the loops $z_i$ with $E_{k,k-q}$ and to the loops $w_i$ with a second copy, selects this coefficient. The order is at most $(5r+1)k+2k+2\cdot20k=(5p+38)k$. Finally, a row scaling removes the sign.
\end{proof}

The following were also checked by exact computation: the identities for $J$, the eight minors of $C$ used in Appendix~\ref{app:selectors},~\eqref{eq:adderper}, Lemma~\ref{lem:pfdet} in orders $2$, $4$ and $6$, part (i) over $\Q$ for $(p,k)=(3,2),(3,1),(5,1)$, and the formula for $H$ with $p=3$ and $k\le2$. The proofs do not depend on these checks.

\section{Proofs of the main theorems}\label{sec:proofs}

In this section, I combine the smooth restriction of Section~\ref{sec:crit}, its realisation as a permanent from Section~\ref{sec:perm}, and the bounds of Section~\ref{sec:numbers}. I begin with the smooth restriction in each characteristic. In odd characteristic $p$ and for $k\ge128$, Corollary~\ref{cor:smooth} gives a form $F=G_D\circ\Lambda$ with $V(F)$ smooth and the parameters~\eqref{eq:params}. By Theorem~\ref{thm:detperm}(ii), $F$ is an affine restriction of $\per_{n'}$ with $n'\le B_pk$. Over $\C$, the construction of~\cite{oai} gives a smooth form of the same degree $d$ in $N_{\C}=\lfloor k^2/100\rfloor$ variables as an affine restriction of $\per_n$ whenever $n\ge11k$. Here $N_{\C}-1\ge k^2/200$. Whereas that construction selects a coefficient inside a permanent and moves the critical set bound to the permanent through characteristic two, the construction in characteristic $p$ uses $G_D$ directly and then Theorem~\ref{thm:detperm}. In particular, no bound over $\C$ is reduced modulo $p$.

\begin{proof}[Proof of Theorem~\ref{thm:main}]
In characteristic $p$, \eqref{eq:params} gives $(N-1)(d-1)\ge k^2(k-1)/63\ge k^3/64$. Hence Theorem~\ref{thm:smooth}(i) gives $\bdc(F)\ge k^3/(256\,e)$ and $\bpc(F)\ge k^3/(512\,e)$. The restrictions are fixed before any representation is chosen, and affine restriction preserves the closures of the classes. Hence these bounds hold for $\per_n$ whenever $F$ is a restriction of $\per_n$.

I take $k=\lfloor n/B_p\rfloor$. If $n\ge128B_p$, then $k\ge128$ and $k\ge n/B_p-1\ge\frac{127}{128}\cdot\frac{n}{B_p}$. Padding with an identity block makes the permanent of order at most $B_pk\le n$ a restriction of $\per_n$. Since $256\cdot(128/127)^3<270$, this gives the two bounds.

Over $\C$, I take $k=\lfloor n/11\rfloor$. If $n\ge1408$, then $k\ge128$ and $k\ge\frac{127}{128}\cdot\frac{n}{11}$. Further, $(N_{\C}-1)(d-1)\ge(k^2/200)(k/4)=k^3/800$, so $\bpc_{\C}(\per_n)\ge k^3/(6400\,e)$. Since $11^3\cdot6400\cdot(128/127)^3<8.8\cdot10^6$, this gives the bound over $\C$. Finally, exact complexity is at least border complexity.
\end{proof}

\begin{proof}[Proof of Theorem~\ref{thm:poly}]
I use the same $F$. Since $d\ge k/3$, the condition $a\le k/6$ gives $a\le d/2$. Hence Theorem~\ref{thm:smooth}(ii) with $c=1/2$ gives $\bdc_a(F)\ge\log(4/3)(N-1)d/(16a)$ and $\bpc_a(F)\ge\log(4/3)(N-1)d/(32a)$. In characteristic $p$, \eqref{eq:params} gives $(N-1)d\ge k^3/63$. Over $\C$, $(N_{\C}-1)d\ge k^3/600$. Affine restriction keeps entry degrees, so the bounds pass to $\per_n$. The ranges in $n$ follow from $k\ge n/12$ and $k\ge n/(2B_p)$.
\end{proof}

I end with the limits of these results. The constants in characteristic $p$ depend on $p$ through $B_p^3$. Hence the bounds improve on the quadratic bound of~\cite{ccl} only when $p$ is small compared with $n^{1/3}$. I claim nothing that is uniform in $p$. I also claim nothing in characteristic two. Further, the cubic order is the limit of this method. A preliminary manuscript by Sheshadri~\cite{sh} shows that arguments which read an invariant of a hypersurface through a corank one kernel system cannot prove more than a constant times $Nd$ for a form of degree $d$ in $N$ variables. For the restrictions used here, $Nd$ is of order $n^3$.

\section{Use of artificial intelligence}\label{sec:ai}

The mathematics was developed with ChatGPT Astra during a discussion on OpenAI’s cubic bound. Some proofs were later refined in conversation with Claude Opus 5.5

\appendix
\section{Proof of the bound for smooth hypersurfaces}\label{app:smooth}

In this appendix, I prove Theorem~\ref{thm:smooth}. I begin by realising a border representation by a single matrix, and I prove two identities for Pfaffians.

\subsection{Limits and Pfaffians}

A border representation is a limit of matrices, and I want a single matrix that represents it. The next lemma provides one, with entries in a field of Laurent series.

\begin{lemma}\label{lem:laurent}
If $F\in\overline{D_{m,a}}$, respectively $F\in\overline{P_{2m,a}}$, then there is a matrix $M(t,x)$, square of order $m$, respectively alternating of order $2m$, with entries in $K((t))[x]$ of $x$-degree at most $a$, such that $P=\det M$, respectively $P=\Pf M$, lies in $K[[t]][x]$ and $P(0,x)=F$.
\end{lemma}

\begin{proof}
The class is the image of a space of matrices under a polynomial map. Hence its closure is irreducible, and it contains a curve through $F$ on which the image is dense. I lift this curve to a curve of matrices, complete the normalisation of the lifted curve, and choose a point over $F$. After replacing the uniformiser by one of its roots if necessary, the completed local ring at this point is $K[[t]]$. The matrix entries then become Laurent series in $t$. The coefficients of the image polynomial are regular at the chosen point, and there they equal the coefficients of $F$.
\end{proof}

I write $\LL$ for an algebraic closure of $K((t))$. Points over the field $K((t))$ are taken over $\LL$, and lengths are compared over $K[[t]]$.

I now turn to Pfaffians. The Pfaffian of an alternating matrix $M$ of order $2m$ is a signed sum over the perfect matchings of $\{1,\dots,2m\}$. A perfect matching splits these indices into $m$ pairs, and its term is the product of the entries $M_{ij}$ over its pairs $\{i,j\}$ with $i<j$. The Pfaffian has integer coefficients, so it makes sense in every characteristic. The matchings that contain a fixed pair $\{i,j\}$ contribute $M_{ij}$ times the Pfaffian of the remaining indices, with one common sign. Therefore, as a polynomial in the independent entries $M_{ij}$ with $i<j$,
\begin{equation}\label{eq:pfder}
\frac{\partial\Pf M}{\partial M_{ij}}=\pm\Pf M_{\ii\jj},
\end{equation}
where $M_{\ii\jj}$ is $M$ with rows and columns $i$ and $j$ removed.

The next identity lets me reduce Pfaffians to determinants. It writes a Pfaffian, up to a factor, as the determinant of a submatrix of one order less.

\begin{lemma}\label{lem:pfdet}
Let $M$ be an alternating matrix of order $2m\ge 2$ over a commutative ring. Let $E$ be its principal submatrix on the indices $3,\dots,2m$, with $\Pf E=1$ when $m=1$, and let $M'$ be the matrix of order $2m-1$ obtained from $M$ by deleting row~$1$ and column~$2$. Then
\[
\det M'=-\Pf(E)\Pf(M).
\]
\end{lemma}

\begin{proof}
Both sides are polynomials with integer coefficients in the entries $M_{ij}$ with $i<j$. Hence I may take these entries to be independent indeterminates over $\Q$. Then $E$ is invertible. Write
\[
M=\begin{pmatrix}A&B\\-B^T&E\end{pmatrix}
\]
with $A$ of order two, and let $B_1$ and $B_2$ be the rows of $B$. Put $\Sigma=A+BE^{-1}B^T$ and $G=\begin{pmatrix}I_2&0\\E^{-1}B^T&I\end{pmatrix}$. Since $E^T=-E$, multiplying out gives $G^TMG=\Sigma\oplus E$. Since $\det G=1$, it follows that $\Pf M=\Sigma_{12}\Pf E$. On the other hand,
\[
M'=\begin{pmatrix}M_{21}&B_2\\-B_1^T&E\end{pmatrix}.
\]
The Schur complement formula therefore gives $\det M'=\det E\cdot(M_{21}+B_2E^{-1}B_1^T)=\det E\cdot\Sigma_{21}$. Finally, $\det E=(\Pf E)^2$ and $\Sigma_{21}=-\Sigma_{12}$. Hence $\det M'=-(\Pf E)^2\Sigma_{12}=-\Pf(E)\Pf(M)$.
\end{proof}

Note that the congruence in this proof also proves the identity $\Pf M=\Pf E\cdot\Sigma_{12}$ in general.

\subsection{The count for smooth hypersurfaces}\label{sec:count}

In this subsection, I prove the count for smooth hypersurfaces that all three results rest on. Throughout, $F\in K[x_1,\dots,x_N]$ is a form of degree $d\ge 2$ with $N\ge 2$, and $V(F)\subset\PP^{N-1}$ is smooth, i.e., the partial derivatives of $F$ have no common zero on $V(F)$.

Two lemmas make this count survive the passage to a limit. The first says that an isolated point of a limit does not disappear before the limit. Instead, its length is shared among isolated points of the family over $K((t))$.

\begin{lemma}\label{lem:spread}
Let $Y$ be a smooth variety of dimension $r$ over an algebraically closed field, and let $\cZ\subset Y\times\Spec K[[t]]$ be cut out locally by $r$ equations. Let $b$ be an isolated point of the special fibre $\cZ_0$, i.e., of the member at $t=0$, and let $\ell(b)$ be its length. Then the generic fibre $\cZ_\eta$, i.e., the member over $K((t))$, has a finite set $S(b)$ of isolated points over $\LL$ whose lengths add up to $\ell(b)$. The sets $S(b)$ for different $b$ are disjoint. Further, a function on $\cZ$ that does not vanish at $b$ vanishes at no point of $S(b)$.
\end{lemma}

\begin{proof}
In local coordinates $u$ on $Y$ at $b$, the completed local ring of $\cZ$ is $B=K[[t,u]]/(f_1,\dots,f_r)$. The ring $B/tB$ is Artinian of length $\ell(b)$. Hence $t,f_1,\dots,f_r$ is a system of parameters of a regular local ring, and therefore a regular sequence in any order. In particular, $t$ is a nonzerodivisor on $B$. Since $B$ is complete in the $t$-adic topology and $B/tB$ is finite, $B$ is a free $K[[t]]$-module of rank $\ell(b)$. Since $b$ is isolated in its fibre and $K[[t]]$ is henselian, $\Spec B$ is an open and closed part of a neighbourhood of $b$ that is finite over $K[[t]]$. Its generic fibre consists of isolated points of $\cZ_\eta$ of total length $\ell(b)$, and I let $S(b)$ be the set of these points. A point of a finite $K[[t]]$-scheme specialises to exactly one closed point, which gives disjointness. Finally, a unit of $B$ stays a unit after $t$ is inverted.
\end{proof}

The second lemma is a form of B\'ezout's theorem that counts only isolated points. It allows the intersection to have extra components, which I cannot rule out.

\begin{lemma}\label{lem:bezout}
Let $Y=\PP^{b_1}\times\dots\times\PP^{b_c}$ over an algebraically closed field, let $r=\dim Y$, and let $D_1,\dots,D_r$ be effective divisors with multidegrees $\delta_1,\dots,\delta_r\in\Z_{\ge0}^c$. Then the isolated points of $D_1\cap\dots\cap D_r$ have total length at most the coefficient $\beta$ of $H_1^{b_1}\cdots H_c^{b_c}$ in $\prod_{i=1}^r(\delta_{i1}H_1+\dots+\delta_{ic}H_c)$. This holds whatever the rest of the intersection looks like.
\end{lemma}

\begin{proof}
Suppose first that $r$ divisors of these multidegrees meet in a finite scheme. Then their equations form a regular sequence at each point, and the local lengths are intersection multiplicities. Hence the total length is the intersection number $\beta$, computed in $\Z[H_1,\dots,H_c]/(H_1^{b_1+1},\dots,H_c^{b_c+1})$. The systems $|\OO_Y(\delta_i)|$ are base point free, so general members $D_i'$ meet in a finite scheme. Let $s_i$ and $s_i'$ be sections defining $D_i$ and $D_i'$, and let $D_i(\lambda)$ be the divisor of $s_i+\lambda s_i'$. These divisors meet in a finite scheme at $\lambda=1$. Hence they also meet in a finite scheme over the generic point of the $\lambda$-line, and there the total length is $\beta$. By Lemma~\ref{lem:spread}, with $\lambda$ in place of $t$, each isolated point of $D_1\cap\dots\cap D_r$ accounts for a separate part of that total, equal to its own length.
\end{proof}

I now return to $F$. The next lemma gives the scheme that I count.

\begin{lemma}\label{lem:polarscheme}
For general $w_1,\dots,w_{N-2}\in K^N$, the scheme $Z=V(F,\partial_{w_1}F,\dots,\partial_{w_{N-2}}F)\subset\PP^{N-1}$ is finite of length $\mu=d(d-1)^{N-2}$.
\end{lemma}

\begin{proof}
Since $V(F)$ is smooth, the polars $\partial_wF$ have no common zero on $V(F)$. Hence they form a base point free system on $V(F)$, and general members cut $V(F)$ properly. B\'ezout's theorem for this proper intersection gives the length.
\end{proof}

Note that $Z$ need not be reduced. In characteristic $p$ it can have fewer points than its length. I never pass from $Z$ to its set of points.

Now suppose that $F$ lies in $\overline{D_{m,a}}$ or in $\overline{P_{2m,a}}$, and let $P=\det M$ or $P=\Pf M$ be as in Lemma~\ref{lem:laurent}. I choose coordinates in which $x_N$ vanishes at no point of $Z$, and I put $\Pi=\{x_N=1\}\subset\A^N$. Let
\[
\cZ=V\bigl(P,\partial_{w_1}P,\dots,\partial_{w_{N-2}}P\bigr)\subset\Pi\times\Spec K[[t]].
\]
Here each derivative is taken on $\A^N$ in the direction $w_j\in K^N$, and then restricted to $\Pi$. The special fibre of $\cZ$ is $Z$.

\begin{lemma}\label{lem:S}
The generic fibre $\cZ_\eta$ has a finite set $S$ of isolated points over $\LL$ whose lengths add up to $\mu$. At each $z\in S$ there is $w_0\in K^N$ with $\partial_{w_0}P(z)\ne0$.
\end{lemma}

\begin{proof}
I apply Lemma~\ref{lem:spread} to the $N-1$ equations on $\Pi$ at every point $b$ of $Z$, and I let $S$ be the union of the sets $S(b)$. Since $V(F)$ is smooth, there is $w_0$ with $\partial_{w_0}F(b)\ne0$. Since $\partial_{w_0}P$ specialises to $\partial_{w_0}F$, the last part of Lemma~\ref{lem:spread} gives $\partial_{w_0}P(z)\ne0$ for $z\in S(b)$.
\end{proof}

The next lemma lifts the points of $S$ to a space that also records kernel vectors. I state it for any square matrix whose determinant is a nonzero multiple of $P$ near $S$. This is all that the Pfaffian case needs.

\begin{lemma}\label{lem:lift}
Let $M'$ be a square matrix of order $h$ whose entries are polynomials in $x$ of degree at most $a$ with coefficients in $K((t))$. Suppose that $\det M'=gP$ for a polynomial $g$ that vanishes at no point of $S$. Then
\begin{equation}\label{eq:beta}
d(d-1)^{N-2}\le\beta_a(h):=[H^{N-1}U^{h-1}V^{h-1}]\,(aH+V)^h(aH+U)^{h-1}\bigl((a-1)H+U+V\bigr)^{N-2}.
\end{equation}
\end{lemma}

\begin{proof}
Let $z\in S$. Since $P(z)=0$, the derivative of $\det M'$ at $z$ in the direction $w_0$ is $g(z)\,\partial_{w_0}P(z)$, which is nonzero. The derivative of a determinant with respect to an entry is a cofactor. Hence some cofactor of $M'(z)$ is nonzero, and $M'(z)$ has rank exactly $h-1$. Its right and left kernels are lines $[v_z]$ and $[u_z]$, and $\adj M'(z)=c\,v_zu_z^T$ with $c\ne0$. A constant change of basis in the column space multiplies $\det M'$ by a nonzero constant. Hence I may assume that the last coordinate of every $v_z$ is nonzero.

I homogenise the entries of $M'$ to degree $a$ with $x_N$, and those of $M'_w=\partial_wM'$ to degree $a-1$. On $Y=\PP^{N-1}\times\PP^{h-1}_u\times\PP^{h-1}_v$ consider
\begin{equation}\label{eq:system}
M'v=0,\qquad (u^TM')_j=0\ \ (j<h),\qquad u^TM'_{w_i}v=0\ \ (1\le i\le N-2).
\end{equation}
These are $h$, $h-1$ and $N-2$ equations of multidegrees $(a,0,1)$, $(a,1,0)$ and $(a-1,1,1)$, so there are $\dim Y$ equations in all. I claim that $\zeta_z=(z,[u_z],[v_z])$ is an isolated point of~\eqref{eq:system}, and that its local scheme is isomorphic to that of $\cZ_\eta$ at $z$. The points $\zeta_z$ are distinct, since their first coordinates are. Hence the claim, Lemma~\ref{lem:S} and Lemma~\ref{lem:bezout} give $\mu\le\beta_a(h)$.

To prove the claim, I work in the local ring of $Y$ at $\zeta_z$, on the charts $x_N=1$, $v_h=1$ and $u_i=1$ for an index $i$ with $(u_z)_i\ne0$. The identity
\[
v_h(u^TM')_h=u^T(M'v)-\sum_{j<h}v_j(u^TM')_j
\]
puts the omitted equation $(u^TM')_h=0$ in the ideal of~\eqref{eq:system}. The $(i,h)$ cofactor of $M'(z)$ is $c\,(v_z)_h(u_z)_i$, which is nonzero. Hence, after moving row $i$ and column $h$ to the front, I can write
\[
M'=\begin{pmatrix}\alpha&\rho\\\gamma&E'\end{pmatrix},\qquad u=\begin{pmatrix}1\\u'\end{pmatrix},\qquad v=\begin{pmatrix}1\\v'\end{pmatrix},
\]
with $E'$ invertible near $z$. The kernel equations say that $v'=-E'^{-1}\gamma$, that $u'^T=-\rho E'^{-1}$, and that $\sigma=0$, where $\sigma=\alpha-\rho E'^{-1}\gamma$. Differentiating gives $\partial_w\sigma=u^T(\partial_wM')v$ for every direction $w$. Therefore, after $u'$ and $v'$ are eliminated, the system on the base is $\sigma=\partial_{w_1}\sigma=\dots=\partial_{w_{N-2}}\sigma=0$. Finally, $\det M'=\pm\det E'\cdot\sigma=gP$, where $\det E'$ and $g$ are units. Hence $\sigma$ is a unit multiple of $P$, and the ideals $(\sigma,\partial_{w_1}\sigma,\dots)$ and $(P,\partial_{w_1}P,\dots)$ coincide.
\end{proof}

\begin{proposition}\label{prop:count}
If $\bdc_a(F)\le m$, then $d(d-1)^{N-2}\le\beta_a(m)$. If $\bpc_a(F)\le m$, then $d(d-1)^{N-2}\le\beta_a(2m-1)$.
\end{proposition}

\begin{proof}
For determinants, I take $M'=M$ and $g=1$ in Lemma~\ref{lem:lift}. For Pfaffians, fix $z\in S$. Since $\partial_{w_0}P(z)\ne0$, the chain rule and~\eqref{eq:pfder} give indices $i<j$ with $\Pf M_{\ii\jj}(z)\ne0$. Replacing $M$ by $Q^TMQ$ for a constant $Q\in GL_{2m}(K)$ multiplies $P$ by $\det Q$. This changes none of the schemes above, and it keeps the entry degrees. For each $z$, the Pfaffian of the principal submatrix of $Q^TM(z)Q$ on the indices $3,\dots,2m$ is a polynomial in $Q$. It does not vanish at a permutation matrix that sends $i$ and $j$ to $1$ and $2$. Since $S$ is finite, one $Q$ makes all of these Pfaffians nonzero. With this $Q$, let $E$ be that principal submatrix and $M'$ the matrix $M$ with row $1$ and column $2$ deleted. By Lemma~\ref{lem:pfdet}, $\det M'=-\Pf(E)\,P$. Hence Lemma~\ref{lem:lift} applies with $h=2m-1$ and $g=-\Pf E$.
\end{proof}

\subsection{Numerical bounds and sharpness}\label{sec:numbers}

Proposition~\ref{prop:count} is an inequality between $d$, $N$ and the order of the matrix. In this subsection, I solve it for the order, which proves Theorem~\ref{thm:smooth}. I then check on Fermat forms that the answer has the right size. For affine linear entries, the coefficient $\beta_1(h)$ is bounded by a single binomial coefficient.

\begin{proof}[Proof of Theorem~\ref{thm:smooth}]
Put $r=N-1$ and $\kappa=2h-1$. All coefficients of the polynomial in~\eqref{eq:beta} are nonnegative. Hence setting $U=V$ and then dehomogenising can only increase the coefficient, and
\[
\beta_a(h)\le[t^r]\,(1+at)^{\kappa}\bigl(2+(a-1)t\bigr)^{r-1}.
\]
For $a=1$ the right hand side is $2^{r-1}\binom{\kappa}{r}$. With $h=m$ for determinants and $h=2m-1$ for Pfaffians, Proposition~\ref{prop:count} gives the two inequalities in (i). Note that $d(d-1)^{r-1}\ge(d-1)^r$ and $\binom{\kappa}{r}\le(e\kappa/r)^r$. Hence the inequalities imply $d-1\le 2^{(r-1)/r}e\kappa/r\le 2e\kappa/r$, i.e., $\kappa\ge r(d-1)/(2e)$. Substituting $\kappa=2m-1$ and $\kappa=4m-3$ gives the bounds on $\bdc$ and $\bpc$.

For (ii), nonnegativity gives $[t^r]f(t)\le f(\tau)/\tau^r$ for every $\tau>0$. I take $\tau=4/(d-a)$, which is positive because $a\le cd<d$. Then $a-1+2/\tau=(d+a-2)/2$, and rearranging gives
\[
\kappa\log\Bigl(1+\frac{4a}{d-a}\Bigr)\ge(N-2)\log\frac{2(d-1)}{d+a-2}+\log\frac{4d}{d-a}.
\]
Since $a\le cd$, it follows that $2(d-1)/(d+a-2)\ge 2/(1+c)$. Further, the last term is at least $\log4$, which is at least $L_c$, and the left side is at most $4a\kappa/((1-c)d)$. Hence $\kappa\ge(N-1)(1-c)L_cd/(4a)$, which is (ii). The bound $\lceil d/a\rceil$ holds because a determinant of order $m$, or a Pfaffian of half order $m$, with entries of degree at most $a$ has degree at most $am$, and so do its limits.
\end{proof}

I now show that these bounds have the right order of growth. To do so, I use the Fermat forms.

\begin{proposition}\label{prop:fermat}
If the characteristic does not divide $d$, the Fermat form $F_{N,d}=x_1^d+\dots+x_N^d$ defines a smooth hypersurface, and both $\bdc_a(F_{N,d})$ and $\bpc_a(F_{N,d})$ are at most $N(\lceil d/a\rceil-1)+1$. So Theorem~\ref{thm:smooth}(i) is tight up to the factor $4e$ for determinants and $8e$ for Pfaffians, and (ii) is tight up to a constant depending on $c$.
\end{proposition}

\begin{proof}
The gradient $(dx_i^{d-1})_i$ vanishes only at the origin, so $V(F_{N,d})$ is smooth. Let $h=\lceil d/a\rceil$. Take $N$ directed paths of $h$ edges from a source to a sink that share no other vertices. Label the $i$th path by $h-1$ copies of $x_i^a$ and one copy of $x_i^{d-a(h-1)}$. Then the sum over the paths of the products of their labels is $F_{N,d}$. Let $A_G$ be the adjacency matrix of this graph in a topological order. Then $\det(I-A_G)=1$, and the (source, sink) entry of $(I-A_G)^{-1}$ is $F_{N,d}$. By the cofactor formula, deleting the row of the sink and the column of the source gives, up to sign, a determinantal representation of order $N(h-1)+1$ with entries of degree at most $a$. Equation~\eqref{eq:pfdet} gives the Pfaffian bound.
\end{proof}

Recall the forms $\sum_ix_i^{p+1}$ from the introduction. More generally, in characteristic $p$ the forms $\sum_ix_i^{p^b+1}$ are smooth, and their general polars are $p^b$th powers of linear forms. Hence their polar scheme has $p^b+1$ points, but its length is $(p^b+1)p^{b(N-2)}$. Note that a count of points instead of lengths would lose the factor $p^{b(N-2)}$ here. Theorem~\ref{thm:smooth} applies to these forms, and by Proposition~\ref{prop:fermat} it is tight on them.

\section{Proof of the critical set bound}\label{sec:crit}

In this appendix, I prove Theorem~\ref{thm:crit} and Corollary~\ref{cor:smooth}. I keep the notation of Section~\ref{sec:ingredients}.

\paragraph{The spectral curve.} For a matrix $X$, let
\[
M=sI_k+tD+uX,\qquad F_X(s,t,u)=\det M,\qquad C_X=V(F_X)\subset\PP^2.
\]
Note that $\det(X+sI+tD)=F_X(s,t,1)$, and that the coefficient of $s^qt^q$ there has degree $d$ in $X$. Hence $G_D(X)=[s^qt^qu^d]F_X$. I need three properties of $C=C_X$. Firstly, $C$ is reduced. Indeed, on the line $u=0$ the form $F_X$ equals $\prod_i(s+t\theta_i)$, a product of distinct linear forms. A repeated factor of $F_X$ would either be divisible by $u$ or restrict to a repeated factor on $u=0$. Hence $F_X$ has no repeated factor. Secondly, every component of $C$ passes through one of the points $P_i=[-\theta_i:1:0]$. At $P_i$ the matrix $M$ is diagonal with exactly one zero entry, so it has rank $k-1$. Thirdly, $F_X$ is monic in $s$. Hence the projection $\pi:C\to\PP^1$, $[s:t:u]\mapsto[t:u]$, is finite and flat of degree $k$, and $\pi_*\OO_C=\bigoplus_{j=0}^{k-1}\OO_{\PP^1}(-j)$. A general fibre of $\pi$ consists of $k$ distinct points, because the discriminant in $s$ does not vanish at $u=0$. I put $\mathcal E=\pi_*\OO_C(k-1)=\bigoplus_j\OO_{\PP^1}(k-1-j)$.

\paragraph{The adjugate scheme.} The adjugate $\adj M$ is the transposed matrix of cofactors, so $M\adj M=\det M\cdot I$. Its entries are forms of degree $k-1$. Let $\mathcal J\subseteq\OO_C$ be the ideal that they generate on $C$, and let $Z=V_C(\mathcal J)$ be the subscheme of $C$ where they all vanish. Since $\adj M$ is nonzero at the points $P_i$, the scheme $Z$ is finite. I keep its full scheme structure.

The next lemma shows that, in degree $k-1$, the forms that vanish on $Z$ are exactly the linear combinations of the adjugate entries.

\begin{lemma}\label{lem:adjspan}
Every form of degree $k-1$ in $s,t,u$ that vanishes on $Z$ is a $K$-linear combination of the entries of $\adj M$.
\end{lemma}

\begin{proof}
Put $A=tD+uX$, and let $\psi:\Mat_k(\OO_{\PP^1})\to\mathcal E$ send $Y$ to $\tr(Y\adj M)$. On $C$, $M\adj M=0$, so $s\adj M=-A\adj M$. Hence, over each chart, the span of the adjugate entries is stable under multiplication by the affine coordinate $s$. Since $s$ generates the coordinate ring of $C$ over the chart, this span is already the ideal that the entries generate. Therefore the image of $\psi$ is $\pi_*(\mathcal J(k-1))$.

Next I study the kernel of $\psi$. The commutators $AY-YA$ with $Y\in\Mat_k(\OO(-1))$ lie in $\ker\psi$, because $\adj M$ commutes with $A$. At the point $[t:u]=[1:0]$ these commutators fill the image of the map $Y\mapsto DY-YD$, which has dimension $k^2-k$. At the same point $\psi$ has rank $k$. Indeed, it sends the diagonal matrix units to the polynomials $\prod_{j\ne i}(y+\theta_j)$, which are linearly independent modulo $\prod_j(y+\theta_j)$. Hence the sheaf $\mathcal I$ of commutators and $\ker\psi$ have the same generic rank, and $\ker\psi/\mathcal I$ is a torsion sheaf. As a quotient of $\OO(-1)^{k^2}$, the sheaf $\mathcal I$ satisfies $H^1(\mathcal I)=0$. Hence $H^1(\ker\psi)=0$ as well. Therefore the constant matrices map onto $H^0(\PP^1,\operatorname{im}\psi)=H^0(C,\mathcal J(k-1))$.

Now let $P$ be a form of degree $k-1$ that vanishes on $Z$. Then $P$ is a section of $\mathcal J(k-1)$, so it agrees on $C$ with a constant combination of adjugate entries. Their difference is a form of degree $k-1$ that is divisible by $F_X$, which has degree $k$. Hence the difference is zero.
\end{proof}

The next lemma is where the condition $X\in\Gamma_D$ enters. It shows that $Z$ is large.

\begin{lemma}\label{lem:Zlarge}
If $X\in\Gamma_D$, then $\length Z\ge T$.
\end{lemma}

\begin{proof}
The derivative of a determinant with respect to an entry is a cofactor. Hence $\partial F_X/\partial X_{ij}=u\,(\adj M)_{ji}$ over any ring. Consider the linear functional $\lambda(P)=[s^qt^qu^c]P$ on forms of degree $k-1$. The equations of $\Gamma_D$ say that $\lambda$ kills every adjugate entry. By Lemma~\ref{lem:adjspan}, $\lambda$ therefore kills every form of degree $k-1$ that vanishes on $Z$.

Now let $P=\sum a_{ij\ell}s^it^ju^\ell$ be a form of degree $q$ that vanishes on $Z$, and fix an exponent $(i,j,\ell)$. The form $s^{q-i}t^{q-j}u^{c-\ell}P$ has degree $2q+c=k-1$, and it vanishes on $Z$. Note that its exponents are nonnegative, since $c\ge q$. Its value under $\lambda$ is $a_{ij\ell}$, which must therefore be zero. Hence $P=0$. In other words, the forms of degree $q$ restrict injectively to $Z$. They form a space of dimension $T$, so $T\le\length Z$. Note that this argument never divides by an integer.
\end{proof}

The scheme $Z$ lies over the singular points of $C$. The next lemma bounds its length by the normalisation defect.

\begin{lemma}\label{lem:delta}
$\length Z\le\delta(C)$.
\end{lemma}

\begin{proof}
Both sides are sums of contributions over the points of $\PP^1$, so I may work over one point. Let $S$ be the local ring of $\PP^1$ at that point. It is a discrete valuation ring with residue field $K$. Over $S$ the curve is $\Spec B$, where $B=S[y]/(f)$ and $f=\det(yI+A_0)$ for some $A_0\in\Mat_k(S)$. Further, $\OO_C/\mathcal J$ becomes $B/J$, where $J$ is the ideal generated by the entries of $\adj(yI+A_0)$. Let $R$ be the normalisation of $B$. It is finite over $S$, since $f$ has distinct roots over the fraction field of $S$. The conductor $\mathfrak c=\{x\in B:xR\subseteq B\}$ is the largest ideal of $R$ that lies in $B$. It suffices to show that $\mathfrak c\subseteq J$ and that $\length_S(B/\mathfrak c)=\length_S(R/B)$.

I begin with the inclusion. Let $\mathcal L$ be the cokernel of $yI+A_0$ on $B^k$. Since $f$ has distinct roots over the fraction field of $S$, $\mathcal L$ is a torsion free $B$-module of rank one. A $B$-linear map $\mathcal L\to B$ is a row $b$ with $b(yI+A_0)\equiv0$ modulo $f$. Lifting $b$ to $S[y]$ and multiplying by the adjugate shows that $b$ is a combination of the rows of $\adj(yI+A_0)$. Hence $J$ is the image of the evaluation map $\Hom_B(\mathcal L,B)\otimes\mathcal L\to B$. Since $K$ is infinite, some $\ell_0\in\mathcal L$ generates $R\mathcal L$ at every maximal ideal of $R$. Then $\mathcal L'=\ell_0^{-1}\mathcal L$ satisfies $B\subseteq\mathcal L'\subseteq R$. For $x\in\mathfrak c$, multiplication by $x$ maps $\mathcal L'\cong\mathcal L$ into $B$ and sends $1$ to $x$. Hence $x\in J$.

For the lengths, let $\lambda_0:B\to S$ read off the coefficient of $y^{k-1}$ in the remainder modulo $f$. In the basis $1,y,\dots,y^{k-1}$, the pairing $(x,x')\mapsto\lambda_0(xx')$ has a Gram matrix that is zero above the antidiagonal and equal to one on it. Hence the pairing identifies $B$ with its dual over $S$. Under this identification, $\Hom_B(R,B)$ becomes the dual of $R$. Further, evaluation at $1$ identifies $\Hom_B(R,B)$ with $\mathfrak c$. Comparing the elementary divisors of $B\subseteq R$ now gives $\length(B/\mathfrak c)=\length(R/B)$.
\end{proof}

The next lemma counts the curves with a large defect. It is the only place where I use an outside result.

\begin{lemma}\label{lem:severi}
In $\PP H^0(\PP^2,\OO(k))$, the reduced curves $C$ with $\delta(C)\ge T$ form a constructible set of dimension at most $k(k+3)/2-T$.
\end{lemma}

\begin{proof}
In every characteristic, the integral plane curves of degree $b$ whose normalisation has genus $g$ form a locally closed set. When this set is nonempty, its dimension is $3b+g-1$~\cite{cht}. Now consider a reduced curve of degree $k$ with components of degrees $b_i$ and normalisation genera $g_i$, for $1\le i\le\alpha$. Comparing the Euler characteristics of $\OO_C$ and $\nu_*\OO_{\widetilde C}$ gives $\delta(C)=\frac{(k-1)(k-2)}{2}-\sum_ig_i+\alpha-1$. By multiplying equations, the reduced curves of a given type form a set of dimension at most $\sum_i(3b_i+g_i-1)=3k+\sum_ig_i-\alpha$. This number equals $k(k+3)/2-\delta(C)$. Since there are finitely many types, the lemma follows.
\end{proof}

Finally, I show that few matrices share the same curve.

\begin{lemma}\label{lem:fibre}
For a fixed curve $C$, the matrices $X$ with $C_X=C$ form a set of dimension at most $N_0=k(k-1)/2$.
\end{lemma}

\begin{proof}
Let $r_X=(1,\dots,1)\adj M$, a row of $k$ forms of degree $k-1$, and let $Z'_X\subset C$ be the subscheme where its entries vanish. I first show that $Z'_X$ has length $N_0$ and that it determines $X$. The row defines a map $\rho_X:\OO_{\PP^1}^k\to\mathcal E$. At $[1:0]$ this map sends the $i$th basis vector to $\prod_{j\ne i}(y+\theta_j)$. This is an isomorphism, and it does not depend on $X$. Hence $\rho_X$ is injective, and its cokernel has length $\chi(\mathcal E)-k=N_0$. As in the proof of Lemma~\ref{lem:adjspan}, $s\,r_X=-r_X(tD+uX)$ on $C$. Therefore the image of $\rho_X$ is $\pi_*$ of the twisted ideal of $Z'_X$, and $\length Z'_X=N_0$. Now suppose that $Z'_X=Z'_{X'}$. Then $\rho_X$ and $\rho_{X'}$ have the same image, so they differ by a constant automorphism of $\OO^k$. This automorphism is the identity, because the two maps agree at $[1:0]$. Hence $r_X=r_{X'}$. Subtracting the relations $s\,r=-r(tD+uX)$ for $X$ and for $X'$ gives $u\,r_X(X-X')=0$. Since $\rho_X$ is injective, $X=X'$.

It remains to show that $\dim\Hilb_{N_0}(C)\le N_0$, where $\Hilb_{N_0}(C)$ is the space of subschemes of $C$ of length $N_0$. Consider a subscheme of $\A^2$ of length $\ell$ supported at one point, together with a basis of its coordinate ring. These data determine two commuting nilpotent $\ell\times\ell$ matrices, namely multiplication by the two coordinates, and a cyclic vector, namely the image of $1$. Conversely, the matrices and the vector determine the subscheme and the basis. Now let $A$ be nilpotent with a centraliser of dimension $j$. Its conjugacy class has dimension $\ell^2-j$. The nilpotent matrices that commute with $A$ satisfy $\det=0$, which is a nontrivial equation on the centraliser. Hence these triples form a set of dimension at most $\ell^2-1+\ell$. The basis accounts for $\ell^2$ of these dimensions, so the subschemes form a family of dimension at most $\ell-1$. Adding one parameter for each support point on $C$ gives $\dim\Hilb_{N_0}(C)\le N_0$. Finally, the pairs $(X,Z')$ with $r_X|_{Z'}=0$ form a closed set. This set maps onto $\{X:C_X=C\}$. It has at most one point over each $Z'$, since $Z'\subseteq Z'_X$ and both have length $N_0$.
\end{proof}

\begin{proof}[Proof of Theorem~\ref{thm:crit}]
By Lemmas~\ref{lem:Zlarge} and~\ref{lem:delta}, the map $X\mapsto C_X$ sends $\Gamma_D$ into the set of reduced curves with $\delta\ge T$. By Lemma~\ref{lem:severi}, this set has dimension at most $k(k+3)/2-T$. By Lemma~\ref{lem:fibre}, the fibres of the map have dimension at most $k(k-1)/2$. Hence $\dim\Gamma_D\le k(k+3)/2-T+k(k-1)/2=k^2+k-T$.
\end{proof}

Note that none of these steps depends on the characteristic. The $k$ distinct points of a general fibre of $\pi$ come from the fixed fibre at $u=0$. The pairing in Lemma~\ref{lem:delta} uses a coefficient and not a trace. Lemma~\ref{lem:fibre} uses a fixed basis at $[1:0]$ and Jordan forms.

\begin{proof}[Proof of Corollary~\ref{cor:smooth}]
Put $v=k^2$ and $\Delta=T-k=N$. The singular set of $V(G_D)\subset\PP^{v-1}$ lies in the projectivisation of $\Gamma_D$. By Theorem~\ref{thm:crit}, it has dimension at most $v-1-\Delta$. In the Grassmannian of $N$ dimensional subspaces, the subspaces through a singular point form a set of dimension at most $N(v-N)+N-1-\Delta=N(v-N)-1$. At a smooth point $[x]$, Euler's identity gives $x\in\ker dG_D(x)$ in every characteristic. Over all smooth points, the subspaces $H$ with $x\in H\subseteq\ker dG_D(x)$ form a set of dimension at most $N(v-N)-1$. Hence a general subspace meets $V(G_D)$ only at smooth points and is tangent nowhere. Therefore $V(F)$ is smooth. Further, $F\ne0$, since otherwise the subspace would be tangent everywhere.

For~\eqref{eq:params}, note that $3q\le k-1$ and $q+1\ge k/3$. Hence $T\ge(q+1)^2/2\ge k^2/18$, and $N-1\ge k^2/18-k-1\ge k^2/21$, where the last step uses $k\ge127$. Further, $d=k-2q\ge(k+2)/3$, which gives the other two inequalities.
\end{proof}

\section{Construction of the selectors}\label{app:selectors}

In this appendix, I build the equality gadget and the selector of Section~\ref{sec:perm}, and I prove Proposition~\ref{prop:selector}.

\subsection{An equality gadget}

Take two marked entries of weights $w$ and $1$ in distinct rows and distinct columns. Delete them, and add three new rows and three new columns carrying
\[
J=\begin{pmatrix}-1&-1&1\\-1&-1&1\\-\frac12&-\frac12&-\frac12\end{pmatrix}.
\]
Join the row of the first entry to new column $1$ with weight $w$, and the row of the second entry to new column $2$ with weight $1$. Join new row $1$ to the column of the first entry with weight $-2$, and new row $2$ to the column of the second entry with weight $1$.

A direct check gives three facts. Firstly, $\per J=1$. Secondly, each of the four $2\times2$ permanents left after deleting one of the first two rows and one of the first two columns of $J$ is $0$. Thirdly, deleting the first two rows and columns of $J$ leaves $J_{33}=-\frac12$. Now fix a matching of everything outside the gadget. The gadget contributes $1$ if neither marked entry is used, and $w\cdot(-2)\cdot(-\frac12)=w$ if both are used. In the two mixed cases it contributes $0$. Hence the gadget forces the two selection bits to agree, and it introduces no other factor. Since the computation is entrywise, several gadgets may share outside rows and columns.

\subsection{The full adder and the circuit}

A full adder takes three bits $a,b,c$ and outputs two bits $s,u$ with $a+b+c=s+2u$, where the sum is taken in the integers. My full adder is a permanent of order eleven, built from
\[
C=\begin{pmatrix}0&0&1&1\\1&0&1&0\\0&0&0&1\\1&1&0&1\end{pmatrix}.
\]
For a set $I$ of the first three columns and a set $O$ of the first three rows, delete them from $C$. Suppose that $O$ is one of $\emptyset$, $\{1\}$, $\{2,3\}$, $\{1,2,3\}$, and that $|I|=|O|$. Then the permanent of what remains is $1$. Read the deleted columns as the input bits $a,b,c$, and the deleted rows as the output bits $(s,u,u)$. Then these eight cases are exactly the solutions of $a+b+c=s+2u$. Table~\ref{tab:adder} lists them.

\begin{table}[h]
\centering
\begin{tabular}{ccccc}
Rows deleted & Columns deleted & $a+b+c$ & $(s,u)$ & Term in \eqref{eq:adderper}\\ \hline
$\emptyset$ & $\emptyset$ & $0$ & $(0,0)$ & $1$\\
$\{1\}$ & one column & $1$ & $(1,0)$ & $sa$, $sb$, $sc$\\
$\{2,3\}$ & two columns & $2$ & $(0,1)$ & $uab$, $uac$, $ubc$\\
$\{1,2,3\}$ & all three & $3$ & $(1,1)$ & $suabc$
\end{tabular}
\caption{The eight cases of the full adder.}\label{tab:adder}
\end{table}

To turn $C$ into a full adder, I put a copy of $C$ and a copy of $C^T$ side by side. For $j=1,2,3$, I join row $j$ of the second copy to column $j$ of the first by an input edge, and row $j$ of the first copy to column $j$ of the second by an output edge. The completions inside the two copies then contribute the square of a permanent of the kind above. I give the input edges the weights $a,b,c$ and output edge $1$ the weight $s$. Finally, I couple output edges $2$ and $3$ by an equality gadget, with $u$ on the first of them. In blocks of sizes $4$, $4$ and $3$, the result is
\begin{equation}\label{eq:adder}
\mathcal A=\begin{pmatrix}C&sE_{11}&U(u)\\\Delta(a,b,c)&C^T&0\\0&V&J\end{pmatrix},
\end{equation}
\[
U(u)=\begin{pmatrix}0&0&0\\u&0&0\\0&1&0\\0&0&0\end{pmatrix},\qquad
V=\begin{pmatrix}0&-2&0&0\\0&0&1&0\\0&0&0&0\end{pmatrix},
\]
where $\Delta(a,b,c)=\mathrm{diag}(a,b,c,0)$ and $E_{11}$ is a $4\times4$ matrix unit. The gadget keeps only the four output sets listed above, and the row and column counts force $|I|=|O|$. Hence
\begin{equation}\label{eq:adderper}
\per\mathcal A=1+s(a+b+c)+u(ab+bc+ca)+suabc.
\end{equation}
Note that the five variables lie in five different rows and five different columns.

Next I describe the circuit. I put the $L$ input bits in the column of weight $1$, and I treat the columns from the lowest weight up. While a column holds at least three live bits, I feed three of them to a full adder. Its sum bit stays in the column, and its carry moves to the next column. Each adder removes one live bit, and every column ends with one or two bits. What remains are two binary numbers of some length $H$. A ripple carry adder of $H$ further full adders turns them into the $H+1$ binary digits of $y_1+\dots+y_L$. If the first phase leaves $B\ge H$ bits, it has used $L-B$ adders. Hence there are at most $L$ adders in all. Note that the circuit is acyclic, and that every intermediate bit is used exactly once.

\begin{proof}[Proof of Proposition~\ref{prop:selector}]
I replace each adder by a copy of $\mathcal A$, and I couple the two ends of every internal wire by an equality gadget. I delete the marked entries that carry a constant input $0$ or an output digit $0$ of $q$. I delete the row and column of each output digit $1$ of $q$, which forces that entry to be used. Finally, I put $z_i$ on the input entry of the bit $y_i$. By~\eqref{eq:adderper} and the gadget, an assignment of all selection bits has weight one if it is consistent at every adder and every wire, and weight zero otherwise. An input vector $y\in\{0,1\}^L$ has exactly one consistent extension. This extension matches the fixed outputs exactly when $y$ has $q$ ones, and it then contributes $\prod_iz_i^{y_i}$. With at most $L$ adders and $2L$ wires, the order is at most $11L+6L$.
\end{proof}

Table~\ref{tab:minors} lists the eight minors of $C$ that the full adder uses, with their permanents.

\begin{table}[h]
\centering
\begin{tabular}{cccc}
Rows deleted & Columns deleted & Remaining matrix & Permanent\\ \hline
$\emptyset$ & $\emptyset$ & $\left(\begin{smallmatrix}0&0&1&1\\1&0&1&0\\0&0&0&1\\1&1&0&1\end{smallmatrix}\right)$ & $1$\\
$\{1\}$ & $\{1\}$ & $\left(\begin{smallmatrix}0&1&0\\0&0&1\\1&0&1\end{smallmatrix}\right)$ & $1$\\
$\{1\}$ & $\{2\}$ & $\left(\begin{smallmatrix}1&1&0\\0&0&1\\1&0&1\end{smallmatrix}\right)$ & $1$\\
$\{1\}$ & $\{3\}$ & $\left(\begin{smallmatrix}1&0&0\\0&0&1\\1&1&1\end{smallmatrix}\right)$ & $1$\\
$\{2,3\}$ & $\{1,2\}$ & $\left(\begin{smallmatrix}1&1\\0&1\end{smallmatrix}\right)$ & $1$\\
$\{2,3\}$ & $\{1,3\}$ & $\left(\begin{smallmatrix}0&1\\1&1\end{smallmatrix}\right)$ & $1$\\
$\{2,3\}$ & $\{2,3\}$ & $\left(\begin{smallmatrix}0&1\\1&1\end{smallmatrix}\right)$ & $1$\\
$\{1,2,3\}$ & $\{1,2,3\}$ & $\left(\begin{smallmatrix}1\end{smallmatrix}\right)$ & $1$\\
\end{tabular}
\caption{The minors of $C$ used by the full adder.}\label{tab:minors}
\end{table}

\end{document}